\documentclass[prl,twocolumn,superscriptaddress,showpacs,preprintnumbers]{revtex4}
\usepackage{graphicx,epsfig}
\begin{document}

\def\beqra{\begin{eqnarray}} \def\eeqra{\end{eqnarray}}
\def\beqast{\begin{eqnarray*}} \def\eeqast{\end{eqnarray*}}
\def\beq{\begin{equation}}      \def\eeq{\end{equation}}
\def\be{\begin{enumerate}}   \def\ee{\end{enumerate}}

\def\gam{\gamma}
\def\Gam{\Gamma}
\def\la{\lambda}
\def\eps{\epsilon}
\def\La{\Lambda}
\def\si{\sigma}
\def\Si{\Sigma}
\def\al{\alpha}
\def\Tha{\Theta}
\def\tha{\theta}
\def\vphi{\varphi}
\def\del{\delta}
\def\Del{\Delta}
\def\ab{\alpha\beta}
\def\om{\omega}
\def\Om{\Omega}
\def\mn{\mu\nu}
\def\mun{^{\mu}{}_{\nu}}
\def\kap{\kappa}
\def\rsi{\rho\sigma}
\def\beal{\beta\alpha}
\def\til{\tilde}
\def\rta{\rightarrow}
\def\eqv{\equiv}
\def\nab{\nabla}
\def\pa{\partial}
\def\sit{\tilde\sigma}
\def\ul{\underline}
\def\indt{\parindent2.5em}
\def\nd{\noindent}
\def\rsi{\rho\sigma}
\def\beal{\beta\alpha}
\def\caa{{\cal A}}
\def\cb{{\cal B}}
\def\cac{{\cal C}}
\def\cd{{\cal D}}
\def\ce{{\cal E}}
\def\cf{{\cal F}}
\def\cg{{\cal G}}
\def\cah{{\cal H}}
\def\ci{{\cal I}}
\def\cj{{\cal{J}}}
\def\ck{{\cal K}}
\def\cl{{\cal L}}
\def\cm{{\cal M}}
\def\cn{{\cal N}}
\def\cO{{\cal O}}
\def\cp{{\cal P}}
\def\car{{\cal R}}
\def\cs{{\cal S}}
\def\ct{{\cal{T}}}
\def\cu{{\cal{U}}}
\def\cv{{\cal{V}}}
\def\cw{{\cal{W}}}
\def\cx{{\cal{X}}}
\def\cy{{\cal{Y}}}
\def\cz{{\cal{Z}}}
\def\asymptotic{{_{\stackrel{\displaystyle\longrightarrow}
{x\rightarrow\pm\infty}}\,\, }} 
\def\asymptext{\raisebox{.6ex}{${_{\stackrel{\displaystyle\longrightarrow}
{x\rightarrow\pm\infty}}\,\, }$}} 
\def\asymptoticp{{_{\stackrel{\displaystyle\longrightarrow}
{x\rightarrow +\infty}}\,\, }} 
\def\asymptoticm{{_{\stackrel{\displaystyle\longrightarrow}
{x\rightarrow -\infty}}\,\, }} 

\def\raisenot{\raise .5mm\hbox{/}}
\def\nota{\ \hbox{{$a$}\kern-.49em\hbox{/}}}
\def\notA{\hbox{{$A$}\kern-.54em\hbox{\raisenot}}}
\def\notb{\ \hbox{{$b$}\kern-.47em\hbox{/}}}
\def\notB{\ \hbox{{$B$}\kern-.60em\hbox{\raisenot}}}
\def\notc{\ \hbox{{$c$}\kern-.45em\hbox{/}}}
\def\notd{\ \hbox{{$d$}\kern-.53em\hbox{/}}}
\def\notbd{\ \hbox{{$D$}\kern-.61em\hbox{\raisenot}}} 
\def\note{\ \hbox{{$e$}\kern-.47em\hbox{/}}}
\def\notk{\ \hbox{{$k$}\kern-.51em\hbox{/}}}
\def\notp{\ \hbox{{$p$}\kern-.43em\hbox{/}}}
\def\notq{\ \hbox{{$q$}\kern-.47em\hbox{/}}}
\def\notW{\ \hbox{{$W$}\kern-.75em\hbox{\raisenot}}}
\def\notz{\ \hbox{{$Z$}\kern-.61em\hbox{\raisenot}}}
\def\notpa{\hbox{{$\partial$}\kern-.54em\hbox{\raisenot}}}

\def\fo{\hbox{{1}\kern-.25em\hbox{l}}}  
\def\rf#1{$^{#1}$}
\def\bx{\Box}
\def\tr{{\rm Tr}}
\def\rmtr{{\rm tr}}
\def\dgg{\dagger}
\def\lag{\langle}
\def\rag{\rangle}
\def\bmid{\big|}
\def\vlap{\overrightarrow{\La p}} 
\def\lrta{\longrightarrow} \def\lrar{\raisebox{.8ex}{$\longrightarrow$}}
\def\ON{{\cal O}(N)}
\def\UN{{\cal U}(N)}
\def\bdPh{\mbox{\boldmath{$\dot{\!\Phi}$}}}
\def\bPh{\mbox{\boldmath{$\Phi$}}}
\def\bPhs{\bPh^2}
\def\sef{S_{eff}[\sigma,\pi]}
\def\sigx{\sigma(x)}
\def\pix{\pi(x)}
\def\bph{\mbox{\boldmath{$\phi$}}}
\def\bphs{\bph^2}
\def\ex{\BM{x}}
\def\exs{\ex^2}
\def\xdot{\dot{\!\ex}}
\def\y{\BM{y}}
\def\ys{\y^2}
\def\ydot{\dot{\!\y}}
\def\pat{\pa_t}
\def\pax{\pa_x}


\title{Stable Fermion Bag Solitons in the Massive Gross-Neveu Model: Inverse 
Scattering Analysis}

\author{Joshua Feinberg}
\affiliation{Department of Physics, University of Haifa at Oranim,
Tivon 36006, Israel.}
\affiliation{Department of Physics, Technion, Haifa 32000, Israel.}
\author{Shlomi Hillel}
\affiliation{Department of Physics, Technion, Haifa 32000, Israel.}

\date{1 September 2005}

\begin{abstract}
Formation of fermion bag solitons is an important paradigm in the theory of 
hadron structure. We study this phenomenon non-perturbatively in the $1+1$
dimensional Massive Gross-Neveu model, in the large $N$ limit. We find, 
applying inverse scattering techniques, that the extremal static bag 
configurations are reflectionless, as in the massless Gross-Neveu model. 
This adds to existing results of variational calculations, which used
reflectionless bag profiles as trial configurations. 
Only reflectionless trial configurations which support a single pair of 
charge-conjugate bound states of the associated Dirac equation were used in 
those calculations, whereas the results in the present paper hold for bag 
configurations which support an arbitrary number of such pairs. We compute 
the masses of these multi-bound state solitons, and prove that only bag 
configurations which bear a single pair of bound states are stable. Each
one of these configurations gives rise to an $O(2N)$ antisymmetric tensor 
multiplet of soliton states, as in the massless Gross-Neveu model. 
\end{abstract}


\maketitle

\section{Introduction}

An important dynamical mechanism, by which fundamental particles
acquire masses, is through interactions with vacuum condensates. Thus, 
a massive particle may carve out around itself a spherical region 
\cite{sphericalbag} or a shell \cite{shellbag} in which the
condensate is suppressed, thus reducing the effective mass of the particle
at the expense of volume and gradient energy associated with the
condensate. This picture has interesting phenomenological consequences
\cite{sphericalbag,mackenzie}.

This dynamical distortion of the homogeneous vacuum condensate configuration,
namely, formation of fermion bag solitons, was demonstrated explicitly 
by Dashen, Hasslacher and Neveu (DHN) \cite{dhn} many years ago, in their
study of semiclassical bound states in the $1+1$ dimensional Gross-Neveu 
(GN) model \cite{gn}.

Fermion bags in the GN model were discussed in the 
literature several other times since the work of DHN, 
using alternative methods \cite{others, papa, josh1}. For a review on 
these and related matters (with an emphasis on the relativistic Hartree-Fock 
approximation) see \cite{thiesreview}. For a more recent review of static 
fermion bags in the GN model (with an emphasis on reflectionless backgrounds 
and supersymmetric quantum mechanics) see \cite{bagreview}. 
The large-$N$ semiclassical DHN spectrum of these fermion bags turns out to 
be essentially correct also for finite $N$, as analysis
of the exact factorizable S-matrix of the GN model reveals \cite{Smatrix}.

A variational calculation of these effects in the $1+1$ dimensional
massive generalization of the Gross-Neveu model, which we will
refer to as MGN, was carried in \cite{FZMGN} a few years ago, and more 
recently in \cite{Thies}. In this paper we study static fermion bags in the 
MGN model using inverse-scattering formalism \cite{inverse}, thus avoiding
the need to choose a trial variational field configuration. The present work 
is thus a natural extension of the inverse-scattering analysis carried out 
by DHN to the massive case.

The MGN model is defined by the action 
\beqra
S&=&\int d^2x\left\{\sum_{a=1}^N\, \bar\psi_a\,\left(i\notpa
-
M\right)\,\psi_a +
\frac{g^2}{2}\; \left(
\sum_{a=1}^N\;\bar\psi_a\,\psi_a\right)^2\right\}\nonumber\\
&:=&\int
d^2x\,\left\{\bar\psi\left[i\notpa-\si\right]\psi-{1\over
2g^2}\left(
\si^2-2M\si\right)\right\}
\label{lagrangian}
\eeqra
describing $N$ self interacting massive Dirac fermions
$\psi_a$ carrying a flavor index $a=1,\ldots,N$, which we
promptly suppress. 

An obvious symmetry of (\ref{lagrangian}) with its $N$ Dirac spinors 
is $U(N)$. Actually, (\ref{lagrangian}) is symmetric under the larger group 
$O (2N)$ \cite{dhn} (see also Section 1 of \cite{bagreview}). 
The fact that the symmetry group of (\ref{lagrangian}) is $O (2N)$ rather than 
$ U(N)$, indicates that it is invariant against charge-conjugation. 
It is easy to see this in a concrete representation for $\gamma$ matrices, 
which we choose as the Majorana representation \cite{dhn, bagreview}  
\beq\label{majorana}
\gam^0=\si_2\;,\; \gam^1=i\si_3\quad {\rm and} \quad 
\gam^5=-\gam^0\gam^1=\si_1\,.
\eeq
(Henceforth, in this paper we will use this representation for  
$\gam$ matrices in all explicit calculations.)

In the representation (\ref{majorana}), charge-conjugation
is realized simply by complex conjugation of the spinor
\beq\label{chargeconjugation}
\psi^c(x) = \psi^*(x)\,.
\eeq
Thus, if $\psi = e^{-i\om t} u(x)$ is an eigenstate of the Dirac equation 
\beq\label{diraceq}
\left[i\notpa-\si (x)\right]\,\psi = 0\,,
\eeq
with energy $\om$ (assuming time independent $\sigx$), then 
$\psi^*(x) = e^{i\om t} u^*(x)$ is an energy eigenstate of (\ref{diraceq}), 
with energy $-\om$. Therefore, the MGN model (\ref{lagrangian}), 
like the GN model, is invariant against charge conjugation, and energy 
eigenstates of (\ref{diraceq}) come in $\pm\om$ pairs.

As usual, the theory (\ref{lagrangian}) can be rewritten with the help of 
the scalar flavor singlet auxiliary field $\si(x)$. Also as usual, we take
the large $N$ limit holding $\lambda\equiv Ng^2$ fixed. Integrating
out the fermions, we obtain the bare effective action
\beq
S[\si] =-{1\over 2g^2}\int\, d^2x
\,\left(\si^2-2M\si\right) -iN\,
\tr~{\rm log}\left(i\notpa-\si\right)\,.
\label{fermout}
\eeq

Noting that $\gam_5 ( i\notpa -\si )= -(i\notpa +\si)\gam_5
$,  we can rewrite the $\tr~{\rm log}(i\notpa -\si )$ as ${1\over 2}
\tr~{\rm log}\left[-(i\notpa -\si)(i\notpa +\si)\right]$. In this paper we 
focus on static soliton configurations. If $\si$ is time independent, the 
latter expression may be further simplified
to $ {T\over 2}\int {d\om \over 2 \pi} [\tr~{\rm log} (h_+-
\om^2)+\tr~{\rm log} (h_-
-\om^2)]$ where 
\beq\label{hpm}
h_{\pm}  \equiv -\pa_x^2 + \si^2 \pm \si'\,,
\eeq
and where $T$ is an infra-red temporal regulator. 

As it turns out, the two Schr\"odinger operators $h_{\pm}$ are
isospectral (see Appendix A of \cite{bagreview} and Section 2 of 
\cite{josh1}) and thus we obtain
\beqra
S[\si] &=&-{1\over 2g^2}\int\, d^2x
\,\left(\si^2-2M\si\right)\nonumber\\ &-& iNT\,
\int\limits_{-\infty}^{\infty} {d\om \over 2 \pi}\tr~{\rm log}(h_- -\om^2)\,.
\label{effective}
\eeqra

In contrast to the standard massless GN model,
the MGN model studied here is not invariant under the $Z_2$
symmetry $\psi\rightarrow \gamma_5 \psi$, $\sigma \rightarrow -\sigma$,
and the physics is correspondingly quite different. As a result of the 
$Z_2$ degeneracy of its vacuum, the GN model contains 
a soliton (the so called CCGZ kink \cite{ccgz, dhn, others, bagreview, josh1})
in which the $\sigma$ field takes on equal and opposite values at 
$x=\pm\infty$. 
The stability of this soliton is obviously guaranteed by topological 
considerations. With any non-zero $M$ the vacuum value of $\sigma$ is 
unique and the CCGZ kink becomes infinitely massive and disappears. If any 
soliton exists at all, its stability has to depend on the energetics of 
trapping fermions.

Let us briefly recall the computation of the unique vacuum of the MGN model.
We shall follow \cite{FZMGN}. For an earlier analysis of the MGN ground state 
(as well as its thremodynamics), see \cite{klimenko}. Setting 
$\si$ to a constant we obtain from (\ref{effective}) the renormalized 
effective potential (per flavor)
\beq\label{veff}
V(\si,\mu) = {\si^2\over 4\pi}~ {\rm log}~ {\si^2\over
e\mu^2} +
{1\over \lambda(\mu)}~\left[{\si^2\over 2} -
M(\mu)\si\right]\,,
\eeq
where $\mu$ is a sliding renormalization scale with
$\lambda(\mu)=Ng^2(\mu)$ and
$M(\mu)$ the running couplings. By equating the coefficient
of $\si^2$ in
two versions of $V$, one defined with $\mu_1$ and the
other with
$\mu_2$,
we find immediately that
\beq\label{scale}
{1\over\lambda(\mu_1)} - {1\over\lambda(\mu_2)} =
{1\over \pi}~{\rm
log}
~{\mu_1\over\mu_2}
\eeq
and thus the coupling $\lambda$ is asymptotically free, just
as in the GN model. Furthermore, by equating the coefficient of $\sigma$
in $V$ we see that the ratio ${M(\mu)\over\lambda(\mu)}$
is a renormalization group invariant. Thus, $M$ and
$\lambda$ have the same scale dependence.

Without loss of generality we assume that $M(\mu)>0$  and
thus the absolute minimum of (\ref{veff}), namely, the vacuum condensate
$m=\langle\si\rangle$, is the unique (and positive) solution of the gap 
equation
\beq\label{gap}
{dV\over d\si}~{\Big|_{\si=m}}= m\left[ {1\over \pi}~{\rm log}
~{m\over\mu} + {1\over \lambda(\mu)}\right] -
{M(\mu)\over\lambda(\mu)} = 0\,.
\eeq
Referring to (\ref{lagrangian}), we see that $m$ is the mass
of the fermion. Using (\ref{scale}), we can re-write the gap equation as
${m\over\lambda(m)}= {M(\mu)\over\lambda(\mu)}$,
which shows manifestly that $m$, an observable physical quantity, is a
renormalization group invariant. This equation also implies that $M(m)=m$, 
which makes sense physically.

Fermion bags correspond to inhomogeneous solutions of the saddle-point 
equation ${\delta S \over \delta \sigma(x,t)}=0$. In particular, static
bags $\sigx$ are the extremal configurations of the energy 
functional (per flavor) 
\beq\label{energydef}\ce [\sigx] = - {S[\sigx]\over NT}\,,
\eeq 
subjected to the boundary condition that $\sigx$ relaxes to its unique vacuum 
expectation value $m$ at $x=\pm\infty$.

The rest of this paper is organized as follows. In the next section we 
express $\ce [\sigx]$ in terms of the scattering-data. DHN have already 
done most of the work for us, save for one crucial piece, specific to the 
MGN model: an expression for $\int_{-\infty}^\infty\, (\sigx -m)\,dx$ in terms
of the scattering data. In this paper we derive this relation. The details of 
this derivation are relegated to the Appendix. We then complete the task of 
writing down $\ce [\sigx]$ in terms of the scattering data. We prove that 
the static extrema of $\ce [\sigx]$ are reflectionless, as in the massless 
GN model  \cite{dhn}, and calculate their masses. Such an extremal 
configuration, considered as a scalar background in the Dirac equation,
typically supports a number $K$ of pairs of charge conjugate bound states at 
energies $\pm\om_n \,(n=1,2,\ldots , K)$ which bind fermions and 
antifermions. Each one of these $\om_n$s depends only on the total number 
$\nu_n$ of fermions and antifermions it binds, thus giving rise to an $O(2N)$ 
rank-$\nu_n$ antisymmetric tensor multiplet of soliton states. As it turns 
out, the mass of such a soliton is a convex function of the $\nu_n$s. 
In Section 3, we invoke the convexity of the soliton's mass formula and prove
that only solitons which support a single pair of bound states 
(i.e., $K=1$) are stable against decaying into lighter solitons. These are
precisely the configurations studied variationally in \cite{FZMGN, Thies}.

\section{Static Configurations and Inverse-Scattering Analysis}

A fermion bag is essentially a trap for fermions. Evidently, an apriori 
specification of a static fermion bag configuration should indicate 
the list of bound states and enumerate the fermions it traps. 
To be specific, we shall evaluate in this section the energy functional 
(\ref{energydef}) of a static configuration $\sigx$, obeying the appropriate 
boundary conditions at spatial infinity, which supports $K$ pairs of bound 
states of the Dirac equation at energies $\pm\om_n$, $n=1,\ldots ,K$ 
(where, of course, $\om_n^2<m^2$). 
The bound states at $\pm\om_n$ are to be considered together, due to 
the charge conjugation invariance of the GN model. Due to Pauli's exclusion 
principle, we can populate each of the bound states $\pm\om_n$ with up to 
$N$ fermions. In such a typical multiparticle state, the 
negative frequency state is populated by $N-h_n$ fermions and the positive 
frequency state contains $p_n$ fermions. In the parlance of Dirac's hole 
theory, this is a many fermion state, with $p_n$ {\em particles} 
and $h_n$ {\em holes} occupying the pair of charge-conjugate bound states 
at energies $\pm\om_n$. We shall refer to the total number of particles and 
antiparticles trapped in the n-th pair of bound states 
\beq\label{nfilling}
\nu_n = p_n+h_n 
\eeq
as the valence, or occupation number of that pair. 

From (\ref{effective}) and (\ref{energydef}) we obtain the bare energy 
functional as 
\beqra\label{stateff}
\ce [\sigx] &=& {1\over 2\lambda} \int\limits_{-\infty}^{\infty}
dx~[V(x) -2M\sigx]\nonumber\\ &-& \int\limits_{-\infty}^{\infty}
{d\om\over 2\pi i}~\tr~{\rm log}~[-\pa_x^2 + V(x) -\om^2]
\eeqra
where 
\beq\label{potential} 
V(x) = \si^2(x) - \si'(x)\,.
\eeq 
(Here we used $\int_{-\infty}^{\infty}dx~\si'(x) =0$ by invoking the
boundary conditions $\sigx\asymptext m$.) 
The expression (\ref{stateff}) is divergent. We regulate it, as usual, 
by subtracting from it the divergent contribution of the vacuum configuration 
$\si^2=m^2$ and by imposing a UV cutoff $\Lambda$ on $\om$. Thus, the 
regulated bare energy functional associated with $\sigx$ is 
\beqra
&&\ce^{reg} [\sigx]  = {1\over 2\lambda}\int\limits_{-\infty}^{\infty} 
dx \,[V(x)-m^2 -2M(\sigx -m)]
\nonumber\\&& + i\int\limits_{-\Lambda}^{\Lambda} {d\om\over 2\pi}\,
\left[\tr\log (h_- - \om^2) -\tr\log (h_{_{VAC}} - \om^2)\right]  
\label{sigenergy}
\eeqra
where 
\beq\label{vacham}
h_{_{VAC}} = -\pax^2 + m^2
\eeq
is the hamiltonian corresponding to the vacuum configuration. 
We are not done yet, as the integrals over $\om$ in (\ref{sigenergy}) 
still diverge logarithmically with the UV cutoff $\Lambda$. However, 
this logarithmically divergent term is precisely the one that should be 
added to the bare $1/\lambda$ in the first term in (\ref{sigenergy}) in order 
to obtain the renormalized coupling appearing in (\ref{scale}) 
\cite{dhn, josh1, bagreview}.

Now that we have written the energy-functional (\ref{sigenergy}) of a
static configuration, or a fermion bag, our next step is to identify those 
fermion bags on which (\ref{sigenergy}) is extremal. 

The energy functional (\ref{sigenergy}) is, in principle, a complicated and 
generally unknown functional of $\sigx$ and of its derivatives. Thus,   
the extremum condition ${\del\ce [\si]\over \del\sigx} =0$, as a functional
equation for $\sigx$, seems intractable. The considerable complexity of the 
functional equations that determine the extremal $\sigx$ configurations is 
the source of all difficulties that arise in any attempt to solve the model 
under consideration. 

DHN found a way around this difficulty in the case of the GN model \cite{dhn}. 
They have used inverse scattering techniques \cite{inverse} to 
express the energy functional $\ce [\si]$  (\ref{sigenergy}) in terms of 
the so-called ``scattering data'' associated with, e.g., the hamiltonian
$h_-$ in (\ref{hpm}) (and thus with $\sigx$), and then solved the extremum
condition on (\ref{sigenergy}) with respect to those data.

The scattering data associated with $h_-$ are \cite{inverse} the 
reflection amplitude $r(k)$ of $h_-$ at momentum $k$, the number $K$ of 
bound states in $h_-$ and their corresponding energies 
$0<\om_n^2\leq m^2\,,(n=1,\ldots, K)$, and also additional $K$ parameters 
$\{c_n\}$, where $c_n$ has to do with the normalization of the $n$th bound 
state wave function $\psi_n$ of $h_-$. More precisely, the $n$th bound 
state wave function, with energy $\om_n^2$, must decay as $\psi_n(x)\sim 
{\rm const.}\exp -\kappa_n x$ as $x\rightarrow\infty$, where  
\beq\label{kappaomega}
0<\kappa_n = \sqrt{m^2 -\om_n^2}\,.
\eeq
If we impose that $\psi_n (x)$ be normalized, this will determine the 
constant coefficient as $c_n$. (With no loss of generality, we may take 
$c_n>0$.) Recall that $r(-k) = r^*(k)$, since the Schr\"odinger potential 
$V(x)$ is real. Thus, only the values of $r(k)$ for $k>0$ enter the scattering 
data. The scattering data are independent variables, which determine $V(x)$ 
uniquely, assuming $V(x)$ belongs to a certain class of potentials which
fall-off fast enough towards infinity.

Since the MGN does not bear topological solitons, neither $h_-$ nor $h_+$ 
can have a normalizable zero energy eigenstate \cite{jackiwrebbi}. 
(See also Section A.1.1 in Appendix A of \cite{bagreview}.) Thus, all the 
$\om_n$ are strictly positive.

We can apply directly the results of DHN in order to write down that part
of (\ref{sigenergy}) which is common to the MGN and GN models, i.e., 
(\ref{sigenergy}) with its term proportional to $M$ removed, in terms of 
the scattering data. In what follows we briefly summarize their results.
(See Sections 2, 3, and Appendix B of \cite{dhn} for details.)

Using the trace identities of the spectral theory of $h_-$, they 
were able to show that 
\beqra\label{traceidentity}
&&I_1[r(k), \{\kappa_n\}] \equiv -{1\over 2\lambda}\int
\limits_{-\infty}^{\infty} dx \,[V(x)-m^2]\nonumber\\
&&= {1\over 2\pi\lambda}\int\limits_{-\infty}^\infty
\,\log \left[1-|r(k)|^2\right]\,dk + {2\over\lambda} \sum_{n=1}^K \kappa_n
\eeqra
This takes care of the first term in (\ref{sigenergy}). Onto the spectral
determinants: These encode the contribution of the Fermi fields to the 
energy of the bag relative to the vacuum. To account for it correctly, we 
should put our field theory in a big spatial box of length $L$ so as to make
the spectrum discrete, which will enable us matching each mode of fermion
fluctuations around the vacuum, with its counterpart, obtained as $\sigx$ is 
turned on adiabatically. We shall take the limit $L\rightarrow\infty$ only 
in the end. Thus, we obtain the Fermi field part of the energy 
(per flavor) as
\beqra\label{fermienergy}
&&\ce_F = i\int\limits_{-\Lambda}^{\Lambda} {d\om\over 2\pi}\,
\left[\tr\log (h_- - \om^2) -\tr\log (h_{_{VAC}} - \om^2)\right]\nonumber\\
&&= -\sum_\alpha\left(\om_\alpha[\sigx] - \om_\alpha^{\rm VAC}\right)
\nonumber\\ 
&&-\sum_{n=1}^K (\om_n - m) + \sum_{n=1}^K {\nu_n\over N}\om_n\,.
\eeqra
The first sum runs over all positive-energy scattering states, where 
$\om_\alpha[\sigx]$ is the energy of the scattering state to which 
the Fermi mode energy $\om_\alpha^{\rm VAC}$ flows to as the vacuum 
configuration $\si=m$ is deformed adiabatically to $\sigx$. 
The second sum in (\ref{fermienergy}) accounts for the first $K$ scattering
states above the threshold $\om=m$ which migrate into the gap to become
the (positive energy) bound states $\om_n$ as $\sigx$ is switched on. 
In the limit $L\rightarrow\infty$ their energies are indistinguishable 
from $m$. Note the minus sign in front of these two sums, as appropriate for
fermion zero-point energy. The last sum in (\ref{fermienergy}) is evidently
the contribution of {\em valence} fermions and antifermions trapped
inside the bag at the bound states $\pm\om_n$. 
By carefully counting scattering modes in the box, DHN arrived at the fairly
standard result 
\beq\label{scaterringstates}
\sum_\alpha\left(\om_\alpha[\sigx] - \om_\alpha^{\rm VAC}\right) = 
-{1\over \pi}\int\limits_m^\infty \delta (\om) d\om
\eeq
where $\delta (\om) $ is the scattering phase-shift. Then, changing to 
momentum space ($k^2 + m^2 = \om^2$), and using a dispersion integral 
representation for $\delta (k)$, DHN derived that 
\beqra\label{phaseshift}
&&I_2[r(k), \{\kappa_n\}] \equiv \int\limits_m^\infty 
\delta (\om) d\om = \nonumber\\ 
&&{1\over{2\pi}}\int\limits_0^\infty\,{k dk\over(k^2 + m^2)^{1/2}}
\,P.P.\int\limits_{-\infty}^\infty\,{\log \left[1-|r(q)|^2\right]\over k-q}
\,dq\nonumber\\ 
&+& 2\int\limits_0^\Lambda
\,{k dk\over(k^2 + m^2)^{1/2}}\,\sum_{n=1}^K \arctan {\kappa_n\over k}\,.
\eeqra
The second integral in (\ref{phaseshift}) can be calculated explicitly. In 
the limit $\Lambda/m >>1$ we obtain 
\beq\label{divergentpiece}
\tilde I_2 = 2\sum_{n=1}^K \left[\kappa_n \log\left({2\Lambda e\over 
m}\right) - {\pi m\over 2} + \om_n \arctan {\om_n\over\kappa_n}\right]\,.
\eeq

DHN's results, Eqs. (\ref{traceidentity}) - (\ref{phaseshift}), correspond
to all terms of (\ref{sigenergy}), save for the term proportional to $M$,
$\int_{-\infty}^\infty \left(\sigx -m\right) \, dx$. This integral cannot be 
expressed in terms of the scattering data based on the trace identities of 
the Schr\"odinger operator $h_-$ discussed in Appendix B 
of \cite{dhn}. Evidently, new analysis is required to obtain its representation
in terms of the scattering data. Happily enough, we were able to obtain such 
a representation. In the Appendix we prove that
\beqra\label{sigmatext}
&&\int\limits_{-\infty}^\infty \left(\sigx -m\right) \, dx =\nonumber\\
&&{1\over{2\pi i}}\int\limits_{-\infty}^\infty\,{\log \left[1-|r(q)|^2\right]
\over im-q}\,dq\nonumber\\&& + 
\sum_{n=1}^K \log\left({m - \kappa_n \over m + \kappa_n}\right)\,.
\eeqra 

Combining (\ref{traceidentity})- (\ref{phaseshift}) and (\ref{sigmatext})
we obtain the desired expression for the energy functional (\ref{sigenergy})
in terms of the scattering data as
\beqra
&&\ce^{reg} [\sigx]  \nonumber\\
&&= -I_1[r(k), \{\kappa_n\}] + 
{1\over \pi}I_2[r(k), \{\kappa_n\}] - I_3[r(k)]\nonumber\\ && + \sum_{n=1}^K 
\left[\left({\nu_n\over N} -1\right)\om_n + m - 
{M\over \lambda}\log\left({m - \kappa_n \over m + \kappa_n}\right) \right]\,,
\label{sigenergyinvscatt}
\eeqra
where 
\beq\label{I3}
I_3[r(k)] = {M\over{2\pi i\lambda}}\int\limits_{-\infty}^\infty\,
{\log \left[1-|r(q)|^2\right]\over im-q}\,dq\,.
\eeq

We shall now extremize (\ref{sigenergyinvscatt}) with respect to the 
scattering data, to obtain the self-consistent static fermion bags in the 
MGN model. Let us vary with respect to $r(k)$ (with $k>0$) first.
As is evident from (\ref{traceidentity}), (\ref{phaseshift}) and 
(\ref{sigmatext}), $\delta \ce^{reg} [\sigx]/\delta r(k) = 
F(k)\, r^*(k)/(1-|r(k)|^2)$, where $F(k)$ is a calculable function, which 
does not vanish identically. (For example, it can be shown that 
${\rm Im}F(k) = - (M/\lambda\pi)k/(m^2 + k^2)$.)
Thus, $r(k) \equiv 0$ is the unique solution of the variational equation 
$\delta \ce^{reg} [\sigx]/\delta r(k) = 0$. Static extremal bags in the 
MGN model are {\em reflectionless}, as their counterparts in the GN model. 

Explicit formulas for reflectionless $\sigx$ configurations with an arbitrary
number $K$ of pairs of bound states are displayed in Appendix B of 
\cite{bagreview}. In particular, the one which supports a single pair of bound 
states at energies $\pm\om_b$ ($\kappa = \sqrt{m^2-\om_b^2}$), the one 
originally discovered by DHN, is
\beqra\label{dhnsoliton}
\sigx = m &+& \kappa \left[ {\rm tanh }~\left( \kappa x -
{1\over 4}~{\rm log}~{m+\kappa\over m-\kappa}\right)\right. \nonumber\\
 &-& \left. {\rm tanh }~\left(\kappa x + {1\over 4}~{\rm log}
~{m+\kappa\over m-\kappa}\right)\right]\,.
\eeqra 

We see that the formidable problem of finding the extremal 
$\sigx$ configurations of the energy functional $\ce [\si]$ 
(\ref{sigenergy}), is reduced to the simpler problem of extremizing an 
ordinary function $\ce(\om_n, c_n) = \ce\left[\si (x; \om_n, c_n)\right]$ 
with respect to the 2$K$ parameters $\{c_n, \om_n\}$ that determine the 
reflectionless background $\sigx$. If we solve this ordinary 
extremum problem, we will be able to calculate the mass of the fermion bag. 
Let us write down this function explicitly:
\beqra
&&\ce^{reg} (\om_n) = \sum_{n=1}^K
\left\{\left( - {2\over\lambda} + {2\over \pi}\log\left({2\Lambda e\over 
m}\right)\right)\kappa_n \right.\nonumber\\
&&+\left(
{\nu_n\over N} -1 + {2\over \pi}\arctan {\om_n\over\kappa_n}\right)\om_n
\nonumber\\
&&\left. - {M\over \lambda}\log\left({m - \kappa_n \over m + \kappa_n}\right)
\right\}\,.
\label{sigenergreflectionless}
\eeqra
The logarithmically divergent term in (\ref{sigenergreflectionless}) should
remind us that this equation is expressed in terms of the bare couplings 
$\lambda$ and $M$. As it turns out, the renormalization procedure of the 
effective potential for the constant condensate, which we reviewed in the 
Introduction, applies also in the case of inhomogeneous background $\sigx$:
The bare coupling $\lambda$ in 
(\ref{sigenergreflectionless}) is related to the renormalized one at 
the energy scale $\mu_2=m$ via an equation identical to (\ref{scale}), namely, 
\beq\label{scalebare}
{1\over\lambda(\Lambda)} - {1\over\lambda(m)} =
{1\over \pi}~{\rm log}~{2\Lambda\over m}\,.
\eeq
(Due to the anisotropic cutoff implied in (\ref{phaseshift}), the cutoff 
scale in (\ref{scalebare}) is $\mu_1 = 2\Lambda$ rather than just $\Lambda$.) 
In addition, since the ratio ${M\over\lambda}$ is a renormalization group 
invariant all the way up to the cutoff scale, we can replace the 
coefficient of the last term in (\ref{sigenergreflectionless}) by the common 
value of that invariant, namely, ${m\over \lambda(m)}$. With the help of 
these two relations, we obtain the renormalized form of 
(\ref{sigenergreflectionless}) as 
\beqra
&&\ce^{ren} (\om_n) = \sum_{n=1}^K
\left\{\left( - {2\over\lambda (m)} + {2\over \pi}\right)\kappa_n \right.
\nonumber\\
&&+\left(
{\nu_n\over N} -1 + {2\over \pi}\arctan {\om_n\over\kappa_n}\right)\om_n
\nonumber\\
&&\left. - {m\over \lambda (m)}\log\left({m - \kappa_n \over m + \kappa_n}
\right)\right\}\,.
\label{sigenergyren}
\eeqra

Next, note that the energy functional $\ce^{ren}[\si]$ evaluated at a 
reflectionless $\si (x;\om_n, c_n)$, is independent of the 
normalization coefficients $c_n$, that do affect the shape of $\sigx$. 
The $c_n$'s are thus {\em flat directions} of $\ce^{ren}[\si]$ in the space
of all reflectionless $\sigx$ configurations. In fact, the $c_n$'s (or more 
precisely, their logarithms) are collective translational coordinates of 
the fermion bag $\sigx$ (see e.g., Appendix B in \cite{bagreview}).
One of these coordinates, corresponds, of course, to global translations of 
the bag as a whole.

Finally, we are left with the task of extremizing 
(\ref{sigenergyren}) with respect to the bound state 
energies $\om_n$ of the reflectionless background $\sigx$. To this end, we 
follow DHN and parametrize these energies as $\om_n = m\cos\,\theta_n$,
with $0 < \theta_n < \pi/2$. (Note that $\theta_n$ cannot attain the 
end-point values of its range: $\om_n=m$ plunges into the continuum, 
and $\om_n=0$ is possible only if $\sigx$ is topologically nontrivial, 
which is not the case in the MGN model.) From (\ref{kappaomega}) we see
that $\kappa_n = m\sin\,\theta_n$. In terms of the angular variables, we may
write (\ref{sigenergyren}) as
\beqra
&&\ce^{ren} (\theta_n) = m\sum_{n=1}^K
\left\{\left( - {2\over\lambda (m)} + {2\over \pi}\right)\sin\theta_n 
\right.\nonumber\\
&&+\left(
{\nu_n\over N} -{2\theta_n\over \pi}\right)\cos\theta_n
\nonumber\\
&&\left. - \gamma \log\left({1 - \sin\theta_n 
\over 1 + \sin\theta_n}\right)
\right\}\,,
\label{sigtheta}
\eeqra
where we have defined the renormalization group invariant  
\beq\label{gamma}
\gamma = {1\over\lambda (m)} = {M(\mu)\over m\lambda(\mu)}\,.
\eeq
Finally, extremizing (\ref{sigtheta}) with respect to $\theta_n$ yields
\beq\label{thetaextremum}
{\partial \ce^{ren}\over \partial\theta_n} = 
2m\left[\left({\theta_n\over\pi} - {\nu_n\over 2N}\right) + \gamma\tan\theta_n
\right]\sin\theta_n = 0\,,
\eeq
thus fixing $\theta_n$ as a function of the {\em filling fraction} 
\beq\label{fillingfraction}
x_n = {\nu_n\over N}\,,\quad 0<x_n<1
\eeq
according to 
\beq\label{thetaext}
{\theta_n\over\pi} + \gamma\tan\theta_n = {x_n\over 2}\,.
\eeq
The fact that the extremal value of $\theta_n$ is determined by the 
total number $\nu_n$ of particles and holes trapped in the bound states 
of the Dirac equation at $\pm\om_n$, and not by the numbers of trapped 
particles and holes separately, is a manifestation of the underlying $O(2N)$ 
symmetry, which treats particles and holes symmetrically. As explained 
in \cite{witten} and in Appendix D of \cite{bagreview}, this fact indicates 
that this pair of bound states gives rise to an $O(2N)$ antisymmetric tensor 
multiplet of rank $\nu_n$ of soliton states. The soliton as a whole is 
therefore the tensor product of all these antisymmetric tensor multiplets. 
The fermion number operator $N_f$ is one of the generators of $O(2N)$.
Its expectation value in the background of the extremal fermion bag in one
of its $O(2N)$ states is (see Section 3 of \cite{bagreview})  
\beq\label{Nffinal}
\langle N_f \rangle = \sum_{n=1\atop \om_n > 0}^K \, (p_n - h_n)\,,
\eeq  
which is simply the sum over the individual valence fermion numbers 
\beq\label{valenceNfn}
N_{f,val}^{(n)} = p_n - h_n 
\eeq
of each of the states in the individual antisymmetric factors. Evidently, 
for each of these antisymmetric representations 
\beq\label{valenceNfnspectrum}
-\nu_n \leq N_{f,val}^{(n)} \leq \nu_n\,,
\eeq
in accordance with charge conjugation invariance.

The left-hand side of (\ref{thetaext}) is a monotonically increasing function.
Therefore, (\ref{thetaext}) has a unique solution in the interval $[0,\pi/2]$. 
This solution is evidently smaller than $\theta_n^{\rm GN} = 
{\pi\nu_n\over 2N}$, the corresponding value of $\theta_n$ in the GN model
for the same occupation number. Thus, the corresponding bound state
energy $\om_n = m\cos\theta_n$ in the MGN model is higher than its
GN counterpart, and thus less bound. 

Substituting the extremal $\theta$'s from (\ref{thetaext}) in 
(\ref{sigtheta}) we find that the mass $\cm$ of our soliton 
(namely, $N\ce^{ren}$ evaluated at the extremal point) is
\beq
\cm (\{\nu_n\}) = Nm\sum_{n=1}^K
\left( {2\over \pi}\sin\theta_n + \gamma \log \,{1 + \sin\theta_n 
\over 1 - \sin\theta_n}\right)\,.
\label{solitonmass}
\eeq
In the case of a single pair of bound states, $K=1$, (\ref{thetaextremum})-
(\ref{thetaext}) and (\ref{solitonmass}) coincide with the corresponding 
results of variational calculations presented in \cite{FZMGN, Thies},
which were based on (\ref{dhnsoliton}) as a trial configuration. In fact, 
it was realized in \cite{Thies} that the trial configuration 
(\ref{dhnsoliton}) is an exact solution of the extremum condition 
${\del\ce [\si]\over \del\sigx} =0$, provided (\ref{thetaext})
is used to fix $\kappa$.  
This choice of trial configuration was very successful indeed!

We should mention that renormalization of the energy functional 
(\ref{sigenergy}) in the background of a generic {\em reflectionless}
$\sigx $ and its extremization with respect to the $\theta_n$s can be
carried out using an alternative method based on the diagonal resolvent
of the Dirac operator \cite{bagreview}, which is basically a generalization
of the calculations in \cite{FZMGN, josh1} to the case of an arbitrary 
number $K$ of pairs of bound states. As it turns out, there are simple 
explicit formulas for the diagonal resolvents of the Dirac operator and 
of $h_-$ in reflectionless $\sigx$ backgrounds, which make these 
computations possible, and lead to (\ref{thetaextremum})- 
(\ref{solitonmass}) \cite{FHunpublished}.

\section{Investigating Stability of Extremal Static Fermion Bags}

The extremal static soliton multiplets which we encountered in the previous 
section, correspond, in the limit $N\rightarrow\infty$, to exact eigenstates 
of the hamiltonian of the MGN model. However, at large but finite $N$, we 
expect some of these states to become unstable and thus to acquire small 
widths, similarly to the behavior of baryons in QCD with 
a large number of colors \cite{wittenlargeN}. The latter are also solitonic 
objects and are analogous to the ``multi-quark'' bound states of the 
MGN and GN models. In particular, Section 9 of \cite{wittenlargeN} offers a 
sketch of the $1/N$ expansion of two-dimensional QCD in the Coulomb gauge, 
both in the baryon and meson sectors, which is similar to the 
corresponding $1/N$ expansion of the GN and MGN models, in the presence
of fermion bags. (This expansion is based on the so-called bilocal 
condensate formalism, which was later developed in \cite{papa,rdmc,bilocal}.)

Furthermore, we can imagine perturbing the MGN action (\ref{lagrangian}) by a 
small $O(2N)$ singlet perturbation (e.g., by adding to (\ref{lagrangian}) a 
term $\Delta S_n = \epsilon\int d^2 x\,\si^{n}$), and ask which of the 
extremal fermion bags of the previous section are stable against such 
perturbations. (The perturbations $S_1$ and $S_2$ correspond merely to a 
redefinition of the bare quantities $M$ and $g^2$, and are thus not 
interesting. The higher perturbations, with $n>2$, are non-renormalizable. 
However, we could think of the resulting perturbed lagrangian as an 
effective one.) Under these circumstances, all possible decay 
channels of a given unstable soliton multiplet must conserve energy, 
momentum and $O(2N)$ quantum numbers. 

It turns out that non-trivial results concerning stability 
may be established without getting into all the details of decomposing
$O(2N)$ representations, by imposing a simple necessary 
condition on the spectrum of the fermion number operator $N_f$ in the 
multiplets involved in a given decay channel. This way of arguing (as 
described in detail below) has led, in the case of the GN model 
(see Section 4 of \cite{bagreview}), to specification of all 
topologically-trivial stable fermion bags, consistent 
with the exact results of \cite{Smatrix}. Thus, it is reasonable to expect
(and conjecture) that applying these stability considerations to the MGN 
model should lead to the correct list of fermion bags in this model 
which remain stable at finite $N$ as well. 
Unfortunately, exact results analogous to \cite{Smatrix} 
and valid for finite $N$, are not available for the MGN model, so this
conjecture has to be verified by explicit calculation of $1/N$ corrections. 
We shall now make these stability considerations explicit.

As we have learned so far, a given static soliton multiplet in our model 
is a direct product of $O(2N)$ antisymmetric tensors. 
The decay products of this soliton also correspond to a direct product 
of antisymmetric tensors. According to (\ref{valenceNfnspectrum}), 
the spectrum  of $N_f$ in an antisymmetric tensor representation is 
symmetric, namely, $-N_f^{max}\leq N_f\leq N_f^{max}$. When 
we compose two such representations $D_1, D_2$, the spectrum of $N_f$ in the 
composite representation $D_1\otimes D_2$, will obviously have the range 
$-N_f^{max}(D_1) -  N_f^{max}(D_2) \leq N_f (D_1\otimes D_2)\leq 
N_f^{max}(D_1) +  N_f^{max}(D_2)$. In particular, each of the possible 
eigenvalues in this range, will appear in at least one irreducible 
representation in the decomposition of $D_1\otimes D_2$. More generally, the 
spectrum of $ N_f (D_1\otimes D_2\cdots \otimes D_L)$ will have the range 
$|N_f (D_1\otimes\cdots \otimes D_L)|\leq N_f^{max}(D_1) + \cdots  
N_f^{max}(D_L)$.

Consider now a decay process, in which a parent static 
soliton, which belongs to a (possibly reducible) representation 
$D_{\rm parent}$, decays into a bunch of other solitons, such that the 
collection of all irreducible representations associated with the decay 
products is $\{D_1, \ldots, D_L\}$ (in which a given irreducible 
representation may occur more than once). By $O(2N)$ symmetry, the 
representation $D_{\rm parent}$ must occur in the decomposition 
of $D_1\otimes D_2\cdots \otimes D_L$. Thus, according to the discussion 
in the previous paragraph, if this decay process is allowed, we must have 
\beq\label{Nfnecessary}
N_f^{max} (D_{\rm parent}) \leq  N_f^{max}(D_1) + \cdots + N_f^{max}(D_L)\,,
\eeq
which is the necessary condition for $O(2N)$ symmetry we sought for. 
(Obviously, similar necessary conditions arise for the other $N-1$  
components of the highest weight vectors of the representations involved.)
The decay process under consideration must respect energy conservation, i.e.,
it must be exothermic. Thus, we supplement (\ref{Nfnecessary}) by 
the requirement 
\beq\label{energynecessary}
\cm_{\rm parent} \geq \sum_{\rm products} \cm_k
\eeq
on the masses  $\cm_i$ of the particles involved.

For each of the static soliton multiplets discussed in the 
previous section, we will scan through all decay channels 
(into static solitons) and check which of these decay channels are 
{\em necessarily} closed, simply by requiring that the two conditions 
(\ref{Nfnecessary}) and (\ref{energynecessary}) be mutually contradictory. 

In order to complete our argument, we must make the plausible 
assumption that for any {\em time dependent} stable soliton with given 
$O(2N)$ quantum numbers (should such a soliton exist), the static soliton with
the same $O(2N)$ quantum numbers is lighter. In the GN this is in fact
true for the known time dependent DHN breathers \cite{dhn}. 

For all the multiplets with a single pair of bound state, $K=1$, we will find 
in this way that {\em all} decay channels are necessarily closed, thus 
establishing their stability. That these are stable multiplets is 
almost obvious to begin with - they are the lightest 
solitons, given their $O(2N)$ quantum numbers. We cannot establish 
in this way that all decay channels are necessarily closed for the 
higher solitons $K>1$, and they are presumably unstable.

Consider the function 
\beq\label{epsilon}
\epsilon (x) = {2\over \pi}\sin\theta (x) + \gamma \log \,{1 + \sin\theta (x)
\over 1 - \sin\theta (x)}\,,\quad 0<x<1\,,
\eeq
where $\theta (x)$ is a solution of (\ref{thetaext}). From (\ref{thetaext})
and (\ref{epsilon}) we obtain that 
\beqra\label{epsilonderivatives}
{d\epsilon (x)\over d x} &=& \cos \theta (x) > 0\nonumber\\
{d^2\epsilon (x)\over d x^2} &=& -{\pi\over 2} {\sin \theta (x)\over 1 + 
\pi\gamma \sec^2 \theta (x)} < 0 \,.
\eeqra
Thus, $\epsilon (x)$ is a monotonically increasing convex function in 
the range of interest, satisfying $\epsilon (x_1 + x_2) 
< \epsilon (x_1)  + \epsilon (x_2)$. 
In terms of (\ref{epsilon}), we may write the soliton mas (\ref{solitonmass})
simply as 
\beq\label{solitonmasx}
\cm (x_1,\cdots,x_n) = Nm\sum_{n=1}^K \epsilon (x_n)\,.
\eeq

Now, we are ready to start the stability analysis. Consider the decaying 
parent soliton to be a static soliton with $K$ pairs of bound states, 
corresponding to the direct product of $K$ antisymmetric 
tensor representations of ranks $\tilde\nu_1, \ldots, \tilde\nu_K$. 
The mass of this soliton is $\cm (\tilde x_1, \ldots, \tilde x_K) = 
Nm\sum_{n=1}^K\,\epsilon (\tilde x_n)$, and according to 
(\ref{valenceNfnspectrum}), the maximal fermion number 
eigenvalue occurring in this representation is $N_f^{max}(D_{\rm parent}) = 
\sum_{n=1}^K\,\tilde\nu_n$. 

Following the strategy which we laid above, we shall now scan through all 
imaginable decay channels of this parent soliton (into final states of purely 
static solitons), and identify those channels which are necessarily closed. 
Thus, assume that the parent soliton under consideration decays into a 
configuration of lighter solitons, with 
quantum numbers of the direct product of $L$ antisymmetric tensor 
representations $\nu_1,\ldots ,\nu_L$. The way in which these $L$ multiplets 
are arranged into extremal fermion bags is of no consequence to our 
discussion. Thus, we discuss all decay channels consistent 
with these quantum numbers in one sweep.

The necessary conditions (\ref{Nfnecessary}) and (\ref{energynecessary}) 
for possible decay imply 
\beq\label{1KintoL}
\sum_{n=1}^K \tilde x_n \leq \sum_{i=1}^L x_i \,,\quad\quad \sum_{n=1}^K 
\epsilon (\tilde x_n) \geq \sum_{i=1}^L \epsilon (x_i)\,,
\eeq
where all $0 < \tilde x_n, x_i < 1$. 
The two pairs of boundary hypersurfaces in (\ref{1KintoL}) are 
\beqra\label{1KintoLsurfaces}
&&\Sigma_1, \tilde\Sigma_1: 
\quad\quad\quad\quad x_1 +\cdots + x_L = \tilde x_1+\cdots + 
\tilde x_K ~~~~~~~~~~~~~\nonumber\\
&&\Sigma_2, \tilde\Sigma_2: 
\quad \epsilon (x_1) +\cdots + \epsilon (x_L) = \epsilon (\tilde x_1) + 
\cdots + \epsilon (\tilde x_K)~~~~~~
\eeqra
where $\Sigma_{1,2}$ are hypersurfaces in $x$-space and 
$\tilde\Sigma_{1,2}$ are the corresponding hypersurfaces in $\tilde x$-space.

Consider the behavior of (\ref{solitonmasx}) over the hyperplane 
\beq\label{constantNfmax}
\tilde\Sigma_{r\alpha}: \quad\quad\quad\quad \tilde x_1+\cdots + 
\tilde x_K = r + \alpha\,,
\eeq
where $0\leq r \leq K$ is an integer and $0\leq \alpha <1$. 
This hyperplane corresponds, of course, to solitons with a fixed value of the 
maximal fermion number $N_f^{max}(D_{\rm parent})$. 
We will now prove that $\cm (x_1,\cdots,x_n)$ attains its absolute 
minimum on $\tilde\Sigma_{r\alpha}$ in the positive orthant, at the 
vertices of the intersection of $\tilde\Sigma_{r\alpha}$ and the hypercube 
$[0,1]^K$, namely, the points
\beq\label{Nfintersectionvertices}
\tilde x_n^{(v)} = \left( \delta_{nn_1} + \delta_{nn_2} + \ldots + 
\delta_{nn_r}\right) + \alpha\delta_{nn_{r+1}}\,,\quad (n=1, \ldots, K)\,,
\eeq
with all possible choices of $r+1$ coordinates $i_1, \ldots, i_{r+1}$ out of 
$K$.

We prove this as follows (see Appendix F of \cite{bagreview}): 
consider a sequence $0\leq\tilde x_1\leq \ldots \leq\tilde x_K\leq 1$, 
subjected to (\ref{constantNfmax}). Assume that for some $i$, 
$0 < \tilde x_i\leq \tilde x_{i+1} < 1 $. We will show that 
there exists another sequence of $\tilde x$'s, with the same sum, but with a 
lower sum of the $\epsilon$s. Thus, let  $\delta > 0$ be chosen such that 
$\tilde x_i - \delta > 0$ and $\tilde x_{i+1} + \delta < 1$, i.e., 
$0\leq\delta\leq\min\{\tilde x_i, 1 - \tilde x_{i+1}\}$. Modify the 
sequence under consideration by replacing $\tilde x_i$ by $\tilde x_i - \
\delta$ and $\tilde x_{i+1}$ by $\tilde x_{i+1} + \delta$, keeping the 
other $K-2$ terms unaltered. The new sequence thus obtained has the same 
sum as the original sequence, and thus defines another point on 
$\tilde\Sigma_{r\alpha}$. We must show that 
$D(\delta) = \cm({\rm original~sequence})  - 
\cm ({\rm new~sequence}) > 0$. Indeed, $D(\delta) =
\epsilon(\tilde x_i) - \epsilon (\tilde x_i - \delta) + \epsilon(\tilde 
x_{i+1}) - \epsilon (\tilde x_{i+1} + \delta)$. Clearly, $D(0) = 0$, 
and also $D\,'(\delta) > 0 $ in the relevant range of $\delta$. Thus, 
$D(\delta)$ increases monotonically with $\delta$, and reaches its maximum 
at $\delta_{\rm max} = \min\{\tilde x_i, 1 - \tilde x_{i+1}\}$, where, 
depending on the initial condition at $\delta = 0$, either 
$\tilde x_i - \delta_{\rm max} =0$ or 
$\tilde x_{i+1} + \delta_{\rm max} =1$. Thus, the sequence of $\tilde x$'s 
constrained to $\tilde\Sigma_{r\alpha}$, which minimizes (\ref{solitonmasx}),
cannot have more than one element in the interior of $[0,1]$. 
Thus, due to (\ref{constantNfmax}), the absolute minimum is the 
sequence in which the $r$ largest $\tilde x$'s are $1$, one $\tilde x$ is 
$\alpha$ and the rest are zero, namely, the vertices 
(\ref{Nfintersectionvertices}). This is just the statement that the mass 
function (\ref{solitonmasx}), being the sum of the convex functions 
$\epsilon (x)$, is convex inside the cube $[0,1]^K$.

Therefore, a parent soliton corresponding to a point in the {\em interior} 
of the intersection of $\tilde\Sigma_{r\alpha}$ and the hypercube $[0,1]^K$,
can decay into a final state with quantum numbers corresponding to the points
(\ref{Nfintersectionvertices}), i.e., $L=K$ and $x_n = \tilde x_n^{(v)}$ in 
(\ref{1KintoL}). In fact, by continuity, such a parent soliton can decay also 
at least into the set of final states contained in small pockets above the 
vertices (\ref{Nfintersectionvertices}), which correspond to $L=K$ and 
$x_n = \tilde x_n^{(v)} + \epsilon_1$ in (\ref{1KintoL}), where 
$\epsilon_1$ is some small calculable quantity, or into final states 
corresponding to $L>K$ in (\ref{1KintoL}), with $x_i = \tilde x_i^{(v)}$ 
for $1\leq i\leq K$ and $x_i = \epsilon_2$ for $K+1\leq i\leq L$, and where 
$\epsilon_2$ is another small calculable quantity.

On the other hand, the parent soliton which corresponds to the vertices 
(\ref{Nfintersectionvertices}) has no open channel to decay through. Hence it
is potentially stable. Indeed, if it could decay through a channel 
corresponding to $x_1, \ldots, x_L$, then, from the requirement that 
these parameters satisfy (\ref{1KintoL}), we would have 
\beqra\label{1KintoLappendix}
\sum_{i=1}^L x_i &>& \alpha + r \nonumber\\
\sum_{i=1}^L \epsilon (x_i) &<& r\epsilon (1) + \epsilon(\alpha)\,.
\eeqra
Define the hyperplane 
\beq\label{hyperplaneappendix}
\Sigma_{r\alpha}: \quad\quad\quad\quad x_1+\cdots + 
x_L = \alpha +  r\,.
\eeq
From the analysis above we know that the absolute minimum of 
$\sum_{i=1}^L \epsilon (x_i) $ over the intersection of 
$\Sigma_{r\alpha}$ and the hypercube $[0,1]^L$ is $r\epsilon (1)+\epsilon 
(\alpha)$. Thus, the points which satisfy the
second inequality in (\ref{1KintoLappendix}) are bounded by 
$\sum_{i=1}^L x_i < \alpha +  r $, in contradiction with the first
inequality in (\ref{1KintoLappendix}). This completes the proof that 
parent solitons which correspond to the vertices 
(\ref{Nfintersectionvertices}) are stable.

Each such vertex represents a soliton which cannot decay through the channel 
under consideration, and is thus potentially stable. More precisely, all 
these vertices correspond to the same soliton, since the coordinates of 
these vertices are just permutations of each other, and thus all of them 
correspond to the same 
set of parameters, in which 
\beqra\label{potentiallystable}
&&r~ {\rm of~the~} \tilde x{\rm 's~ are~ degenerate~ and~ equal~ to~} 1
\nonumber\\
&&{\rm one~ of~ the~} \tilde x{\rm 's~ is~equal~ to~} \alpha\,,~ {\rm and}
\nonumber\\ 
&&{\rm the~remaining~} K-(r+1)~ \tilde x{\rm 's~ are~ null.}
\eeqra 

Does such a soliton exist? To answer this question let us recall a few 
basic facts: The parent soliton under discussion is topologically trivial. As
such, it must bind fermions to be stabilized, and none of its bound state
energies may vanish. Thus, all the ranks occurring in it must satisfy 
$0 < \tilde\nu_n < N$. Finally, note, that due to the elementary fact, that 
the spectrum of the one dimensional Schr\"odinger operator $h_-$ cannot be 
degenerate, all the $\om_n$'s must be different from each other, and so must
be the $\tilde\nu_n$s. Thus, the only physically realizable parent solitons, 
which are {\em necessarily} stable against the decay channel in question, 
correspond to $r=0$ and $K=r+1 = 1$. These are, of course, the static 
solitons studied in \cite{FZMGN, Thies}.

A corollary of our analysis so far is that these $K=1$ solitons are stable 
against decaying into $L_f$ free fermions or antifermions 
(i.e., $L_f$ fundamental $O(2N)$ representations) plus $L-L_f$ solitons 
corresponding to higher antisymmetric tensor representations. 
In particular, it is stable against complete evaporation into free fermions. 
(Strictly speaking, this argument is valid only for values of $L_f$ which 
are a finite fraction of $N$, since our mass formula (\ref{solitonmasx}) 
is the leading order in the $1/N$ expansion, while removing a finite number 
of particles from the parent soliton is a perturbation of the order $1/N$ 
relative to (\ref{solitonmasx}).) 
Stability of the $K=1$ solitons was studied in \cite{FZMGN} in detail. 
In particular, the regimes $\gamma <<1$ and $\gamma >>1$ were investigated. 
It was noted in \cite{FZMGN} that the mass of the most stable
soliton, obtained as the filling fraction $x\rightarrow 1$, is nonanalytic
around $\gamma =0$ (with leading nonanalytic behavior $\gamma \log \gamma$). 
It was shown in \cite{FZMGN} that this nonanalyticity was related to the 
enhanced $Z_2$ symmetry of the GN model at $\gamma =0$. The opposite regime, 
$\gamma = M/\lambda m >>1$ may be attained by making the four-fermi 
interactions weak. In this regime the theory should be described in terms of 
quasi-free massive fermions of mass $m$. We thus expect that the binding 
energy of bags will tend to zero as $\gamma \rightarrow\infty$, which was 
indeed verified in \cite{FZMGN}.

\section{Appendix: Derivation of (\ref{sigmatext})}

Consider Schr\"odinger equation  
\beq\label{schrodingereq}
\left(-\pax^2 + V(x) -k^2 -m^2\right)\Psi(x,k) = 0
\eeq
with $V(x)$ given by (\ref{potential}). Due to the boundary conditions
on $\sigx$ at spatial infinity, $V(x) \asymptext m^2$. 
Let $\phi(x,k)$ be its scattering solution
\beqra\label{phi}
t(k)\phi(x,k) &=& t(k)e^{ikx} + o(1)\,\quad\quad\quad\quad ~~ 
x\rightarrow +\infty\nonumber\\
t(k)\phi(x,k) &=& e^{ikx} + r(k)e^{-ikx} + o(1)\,\quad
x\rightarrow -\infty
\eeqra
where $t(k)$ and r(k) are, respectively, the transmission and reflection
amplitudes. As a consequence of analyticity of these amplitudes, the 
transmission amplitude $t(k)$ is completely determined on the real $k$ axis 
by $r(k)$ as 
\cite{inverse}
\beqra\label{transmission}
&&t(k) = \sqrt{1-|r(k)|^2}\,
\left(\prod_{l=1}^K {k+i\kappa_l\over k-i\kappa_l}\right)\cdot\nonumber\\
&&\exp\left({1\over 2\pi i}{\rm P.P.}
\int\limits_{-\infty}^\infty\,{\log\,\left[1-|r(q)|^2\right]\over q-k}dq
\right) = \nonumber\\
&&\left(\prod_{l=1}^K {k+i\kappa_l\over k-i\kappa_l}\right)\cdot\nonumber\\
&&\exp\left({1\over 2\pi i}\,
\int\limits_{-\infty}^\infty\,{\log\,\left[1-|r(q)|^2\right]\over q-k 
-i\epsilon}dq\right)\,.
\eeqra
Evidently, the last expression can be extended to the upper-half 
complex $k$-plane (as usual, $\epsilon \rightarrow 0+$),
displaying the bound states poles of $t(k)$ in the upper half-plane 
explicitly. The function 
\beq\label{zeroomega}
\Psi_0(x) = c\, \exp -\int\limits_0^x \si(y)dy
\eeq
solves (\ref{schrodingereq}) for $\om=0$, i.e., for $k=im$. Since 
$\sigx\asymptext m$, it is not normalizable . Moreover, we can always choose 
the constant $c$ such that $\Psi_0(x)\asymptoticp  e^{-mx} = e^{i(im)x}$. 
Thus, with this choice of $c$,
\beq\label{psiphi}
\Psi_0(x) = \phi(x,im)\,.
\eeq
Therefore, from (\ref{phi}), we obtain that
\beq\label{asymptoticphi}
{\phi(L,im)\over \phi(-L,im)}{_{\stackrel{\displaystyle\longrightarrow}
{L\rightarrow\infty}}\,\, }\,t(im)e^{-2mL}\,.
\eeq
With the help of (\ref{zeroomega}) we see that (\ref{asymptoticphi})
is equivalent to 
\beq\label{sigmaint1}
\int\limits_{-\infty}^\infty \left(\sigx -m\right) \, dx = -\log t(im)\,.
\eeq
Thus, finally, 
\beqra\label{sigmaint}
&&\int\limits_{-\infty}^\infty \left(\sigx -m\right) \, dx =\nonumber\\
&&{1\over{2\pi i}}\int\limits_{-\infty}^\infty\,{\log \left[1-|r(q)|^2\right]
\over im-q}\,dq\nonumber\\&& + 
\sum_{n=1}^K \log\left({m - \kappa_n \over m + \kappa_n}\right)
\eeqra 
from (\ref{transmission}), thus proving (\ref{sigmatext}). 
Due to the fact that on the real $q$-axis $r(-q) = r^*(q)$ and $|t(q)|^2 = 
1-|r(q)|^2 \leq 1$, we can rewrite (\ref{sigmaint}) as 
\beqra\label{sigmaintreal}
&&\int\limits_{-\infty}^\infty \left(\sigx -m\right) \, dx =\nonumber\\
&&-{m\over{\pi}}\int\limits_0^\infty\,{\log |t(q)|^2
\over q^2 + m^2}\,dq + \sum_{n=1}^K 
\log\left({m - \kappa_n \over m + \kappa_n}\right)\,.
\eeqra

\begin{acknowledgments}
JF thanks Michael Thies for useful correspondence. 
\end{acknowledgments}

\typeout{References}

\end{document}